\documentclass{article}

\usepackage{wrapfig}
\usepackage{amsmath}
\usepackage{amssymb}
\usepackage{fullpage}
\usepackage{enumerate}
\usepackage{graphicx}
\usepackage{subcaption}
\usepackage{ifthen}
\usepackage{xcolor}
\usepackage{authblk}

\usepackage[nottoc]{tocbibind}

\usepackage[a4paper,top=3cm,bottom=2cm,left=3cm,right=3cm,marginparwidth=1.75cm]{geometry}
\title{Fine-tuning and the doublet-triplet splitting problem in the minimal $SU(5)$ GUT}

\author[1]{Dani\"el Boer\thanks{d.boer@rug.nl}}

\author[1]{Ruud Peeters\thanks{r.j.c.peeters@rug.nl}}

\affil[1]{Van Swinderen Institute for Particle Physics and Gravity, University of Groningen, Nijenborgh 4, 9747 AG Groningen, The Netherlands}
\date{\today}

\newcommand{\order}[1]{$\mathcal{O}(#1)$}
\newcommand{\nn}{\nonumber}

\usepackage[normalem]{ulem} % \sout{old text} for strikeout
\renewcommand\sout{\bgroup \color[rgb]{0.55,0.00,0.99} \ULdepth=-.5ex \ULset}

\begin{document}
\maketitle
\begin{abstract}
In this paper we analyse the doublet-triplet splitting problem in the minimal non-super-symmetric $SU(5)$ GUT. We take into account the full symmetry breaking pattern with both high scale $SU(5)$ breaking and electroweak symmetry breaking. Our analysis shows that the only phenomenologically acceptable model has three vevs, with a strong hierarchy determined by the minimization conditions. The amount of fine-tuning in the model is then numerically evaluated by looking at the effect of variation of input parameters on both the minimization conditions and the bosonic masses. Regarding the vevs as output parameters, a large amount of fine-tuning is required in this scenario, which is an expression of the doublet-triplet splitting problem. We show that this problem is more general, since a model with coupled scalar sectors will in general never realise a hierarchy in vevs. To avoid these problems we advocate imposing the desired hierarchy in vevs as part of the theory. We argue for this viewpoint because the $SU(5)$ breaking and electroweak symmetry breaking need to be adjusted to each other anyway and cannot be regarded as independent mechanisms. We suggest that not only the symmetry breaking pattern needs to be imposed, but also the scales at which the breakings happen. We show quantitatively that the generic theory with hierarchy imposed does not require any fine-tuning of the free parameters which can all be natural and perturbative as desired.
\end{abstract}

\section{Introduction}
In the search for Beyond the Standard Model (BSM) physics, Grand Unified Theories (GUTs) have been a prime candidate for a long time. These theories can provide a common origin for the different forces we observe at low energy, and give an explanation for the quantization of charge. The first GUT put forward was based on $SU(5)$ symmetry \cite{Georgi:1974sy}. It is currently considered ruled out, because it leads to a too short lifetime of the proton compared to its experimental lower limit. Besides this lack of phenomenological viability, it is also considered to suffer from a theoretical problem common to many supersymmetric and non-supersymmetric GUTs: the doublet-triplet splitting problem (DTSP) \cite{Mohapatra:1997sp, Dimopoulos:1981zb}. This entails the problem of producing {\it without fine-tuning} masses of very different scales for the triplet and the doublet part of the scalar sectors. Such fine-tuning seems required already at leading order (classically). For specific GUTs there exist solutions to the DTSP, e.g.\ the Dimopoulos-Wilczek solution for $SO(10)$ \cite{Dimopoulos:1981dw}, the missing partner mechanism \cite{Dimopoulos:1981zu}, the sliding singlet mechanism \cite{Witten:1981kv} and GIFT \cite{Inoue:1985cw, Anselm:1986um}. These solutions often rely on including additional representations in the model. Here we will reanalyse the problem in the minimal non-supersymmetric $SU(5)$ GUT and address the problem of fine-tuning quantatitively by investigating both the equations that define the minimum of the potential and the masses of the scalars and gauge bosons. We first investigate which symmetry breaking pattern is required for a phenomenologically viable theory. Then we determine the fine-tuning, taking the vacuum expectation values (vevs) as the output parameters of the model. After concluding that the model indeed needs a large amount of fine-tuning, we analyse a similar situation in the two Higgs doublet model (2HDM) for which one can analyse the problem analytically. This analysis shows that it is actually a general problem to obtain a hierarchy in vevs in a model with coupled scalar sectors. We then argue for a different viewpoint where the vevs have to be considered input parameters of the theory. 
This amounts to not only imposing the orientation of the various vevs, as is done in the standard viewpoint, but also imposing at which scales a specific symmetry breaking occurs. We evaluate the fine-tuning using this approach, and find that the general theory with a hierarchy in vevs will not require fine-tuning of the remaining free parameters. It must be pointed out that all this concerns leading order and does not pertain to the standard hierarchy problem arising from loop corrections to the scalar boson masses nor to the gauge hierarchy problem \cite{Gildener:1976ai,Natale:1982mt,Mohapatra}.

\section{The doublet-triplet splitting problem}
In order to discuss the doublet-triplet splitting problem (DTSP), we will start by recalling the minimal Higgs potential of $SU(5)$. This potential contains two fields: the $\Phi$ field in the adjoint (\textbf{24}) representation, and the $H$ field in the vector (\textbf{5}) representation \cite{Mohapatra,Sherry:1979sz}. In terms of real scalar fields we write:
\begin{equation}
H = \frac{1}{\sqrt{2}}\begin{pmatrix}
H_1 + i H_2 \\
H_3 + i H_4 \\
H_5 + i H_6 \\
H_7 + i H_8 \\
H_9 + i H_{10}
\end{pmatrix},
\quad
\Phi = \sum_{i=1}^{24}\frac{\lambda_i}{2}\phi_i,
\end{equation}
with $H_i$ and $\phi_i$ real scalar fields, and $\lambda_i$ the traceless generators of $SU(5)$ satisfying Tr$\lambda_i\lambda_j = 2 \delta_{ij}$.

We will impose a $\mathbb{Z}_2$ symmetry that transforms $\Phi \rightarrow -\Phi$. This will be justified later. When imposing this symmetry, the potential is given by:
\begin{align}
\begin{split}
V(H,\Phi) = &-\frac{\nu^2}{2} H^{\dagger}H + \frac{\lambda}{4}(H^{\dagger}H)^2 - \frac{\mu^2}{2}\text{Tr }\Phi^2 + \frac{a}{4}(\text{Tr }\Phi^2)^2 + \frac{b}{2} \text{Tr }\Phi^4 \\ &+  \alpha H^{\dagger}H\text{Tr }\Phi^2 + \beta H^{\dagger}\Phi^2H
\end{split}
\label{eq:potential}
\end{align}
This $SU(5)$ symmetric potential is broken to the Standard Model gauge group by setting the vacuum expectation value of the $\Phi$ field: $\langle \Phi \rangle = \text{diag}(v,v,v,-\frac{3}{2}v, -\frac{3}{2}v)$, where the factors $-\frac{3}{2}$ are necessary to ensure the tracelessness of $\Phi$. This form can be achieved by setting $\langle\phi_{24}\rangle = \sqrt{15}v$. If we consider the potential with this vev, we can write down one condition defining the minimum of the potential (a minimum equation):
\begin{equation}
\mu^2 = \frac{15a +7b}{2}v^2
\label{eq:GUTmin}
\end{equation}
Now we can look at the masses of the doublet and triplet components of the $H$ field. If we look at the potential at the minimum of $\Phi$, we can write down the purely $H$ dependent part of the potential, and write it in terms of the doublet $\vec{H}_2$ and the triplet $\vec{H}_3$:
\begin{align}
\begin{split}
V(\vec{H}_2,\vec{H}_3) = &\left[-\frac{\nu^2}{2}+\frac{15\alpha}{2}v^2 + \beta v^2\right]\vec{H}_3^{\dagger}\vec{H}_3 + \left[-\frac{\nu^2}{2}+\frac{15\alpha}{2}v^2 + \frac{9\beta}{4} v^2\right]\vec{H}_2^{\dagger}\vec{H}_2 \\ &+ \frac{\lambda}{4}\left[(\vec{H}_3^{\dagger}\vec{H}_3)^2 + (\vec{H}_2^{\dagger}\vec{H}_2)^2 + 2\vec{H}_3^{\dagger}\vec{H}_3\vec{H}_2^{\dagger}\vec{H}_2\right] 
\end{split}
\end{align}
From this we can read of the masses of the doublet and triplet:
\begin{align}
m^2_{\vec{H}_2} &= -\frac{\nu^2}{2}+\frac{15 \alpha}{2}v^2 + \frac{9\beta}{2} v^2\\
m^2_{\vec{H}_3} &= -\frac{\nu^2}{2}+\frac{15 \alpha}{2}v^2 + \beta v^2
\end{align}

In order to discuss the amount of fine-tuning in this model, we need to write the masses in terms of the input parameters. These are usually considered to be the mass parameters $\mu^2$ and $\nu^2$ and the couplings in the Lagrangian. The vev is generally not considered an input parameter, as it arises when the energy of the system is lowered and is determined by the other parameters using the minimum equation (Equation \ref{eq:GUTmin}). When replacing the vev we find:
\begin{align}
m^2_{\vec{H}_2} &= -\frac{\nu^2}{2}+\frac{15 \alpha + 9\beta}{15a + 7 b}\mu^2 \label{eq:mass2}\\
m^2_{\vec{H}_3} &= -\frac{\nu^2}{2}+\frac{15 \alpha + 2\beta}{15a + 7b}\mu^2 \label{eq:mass3}
\end{align}
The mass parameters $\mu^2$ and $\nu^2$ originate from the $\Phi$ and $H$ sector respectively. When we consider these sectors separately there is no reason that the mass parameters are of the same order of magnitude. We would expect them to be quite different. The interactions between the $\Phi$ and $H$ sector do not change this reasoning, at least at tree-level. Applying this to Eqs.\ (\ref{eq:mass2}) and (\ref{eq:mass3}), we see that both masses will in general be of the order of the highest scale. This also implies that these masses are of the same order of magnitude.

However, phenomenological constraints require a hierarchy between the triplet and doublet mass. The triplet can mediate proton decay, so its mass should be at least \order{10^{14}} GeV \cite{Mohapatra}, which is the typical scale at which the coupling constants become of the same order and could unify. When we assume that the couplings are all \order{1}, i.e.\ natural and perturbative, and there are no cancellations between $a$ and $b$, this means that at least one of $\nu^2$ and $\mu^2$ has to be of the order of this high scale. But the doublet should behave like the Standard Model Higgs doublet, giving masses of \order{100} GeV to the electroweak gauge bosons and the Higgs boson. Therefore we expect $m^2_{\vec{H}_2} \approx 100$ GeV. Now assuming that there is no large cancellation between $\alpha$ and $\beta$, it means that the large contribution we needed for the heavy triplet now has to cancel with the other contribution to a very high degree in order to result in a doublet that is much lighter than the triplet. This is the essence of the doublet-triplet splitting problem. So in order to have the hierarchy between the triplet and doublet mass, one  either needs a cancellation between $\alpha$ and $\beta$ if $\mu = {\cal O}(v) \gg \nu$, or there is a cancellation between the $\mu^2$ and $\nu^2$ term, in which case both have to be of the high scale $v$. In either case there is fine-tuning required at tree-level. In conclusion, the generic case will not lead to the desired hierarchy.

The discussion so far does not take into account electroweak symmetry breaking (EWSB), but that is expected to play a role as it must be compatible with the $SU(5)$ symmetry breaking and allow for a phenomenologically viable scenario. In the next section we will investigate the role of EWSB, and see which possibilities there are to realise EWSB in a satisfactory way. That discussion will prompt us to reconsider the standard viewpoint of what should be considered input parameters at high scales.

\section{Including electroweak symmetry breaking}
\label{sec:EWSB}
The most straightforward implementation of EWSB is by giving the $H$ field a vev of the form: $\langle H \rangle = \frac{1}{\sqrt{2}}(0,0,0,0,v_0)^T$, in addition to the previously defined vev of $\Phi$. In that case we obtain three minimum equations:
\begin{align}
\nu^2 &= (15\alpha + \frac{9}{2}\beta)v^2 + \frac{1}{2}\lambda v_0^2 \\
0 &= -\frac{3}{2}\beta v v_0  \label{eq:betazero}\\
\mu^2 &= (\frac{15}{2}a + \frac{7}{2}b)v^2 + (\alpha+\frac{3}{10}\beta)v_0^2
\end{align}
So this implies the additional constraint that $\beta = 0$. If we continue analyzing the model, and calculate the masses of the scalars, we see that we get 21 massless modes, whereas one only expects 15 Goldstone bosons (12 from the breaking of $SU(5)$ and 3 from EWSB). The six additional massless scalars arise since the potential has a larger symmetry when we set $\beta = 0$. In that case the $H$ and $\Phi$ sector no longer mix, and there is an additional global $SU(5)$ symmetry for the $H$ field. This symmetry remains unbroken when breaking the local $SU(5)$ symmetry. But when $H$ gets a vev, it breaks from a global $SU(5)$ to a global $SU(4)$. This results in 9 massless modes. Three of these are eaten by the electroweak gauge bosons, the other six remain present in the theory as massless modes. This is phenomenologically not an acceptable scenario.

Besides phenomenological considerations, another reason to discard this scenario is that the input parameter $\beta$ needs to have a specific value to ensure a certain low energy outcome. The constraint $\beta = 0$ only arises from the electroweak symmetry breaking, it is not present when only breaking the $SU(5)$ symmetry. So the theory at the GUT scale has to ``know" already about the low energy physics in order to satisfy the minimum equations and have $\beta = 0$.

Beyond leading order gauge boson loops will induce a $H^{\dagger}\Phi^2H$ term, leading to a non-zero value for $\beta$. This can be seen from the $SU(5)$ effective potential \cite{Georgalas:1982mc}\footnote{Note there must be an error in Eq.\ 2.4 of \cite{Georgalas:1982mc} because there is a non-zero contribution to the $H^{\dagger}\Phi^2H$ term from Higgs loops even when $\beta = 0$. This is impossible, since there are no terms present in the Higgs potential that mix components of $H$. It was checked that the gauge boson loop really does give rise to an effective $\beta$ parameter}. Hence, beyond leading order the above argument may not apply, because the Goldstone bosons may be absent and the minimum equations will be different, but as mentioned in the introduction we restrict to a discussion of the DTSP at tree level. Beyond leading order one has to deal with the hierarchy problems in addition.

The restriction on $\beta$ can be lifted at leading order already by realizing that the $\Phi$ field can also contribute to EWSB, see e.g.\ \cite{Buras:1977yy}. This is done by adding a vev $v_1$, which leads to $\langle \Phi \rangle = \text{diag}(v, v, v, -\frac{3}{2}v -\frac{1}{2}v_1, -\frac{3}{2}v + \frac{1}{2}v_1)$. The $H$ field has the same vev as before: $\langle H \rangle = \frac{1}{\sqrt{2}}(0,0,0,0,v_0)^T$. When we write down the minimum equations in this case, we obtain:
\begin{align}
\nu^2 &= (15\alpha + \frac{9}{2}\beta)v^2 - 3\beta v v_1 + (\alpha + \frac{1}{2}\beta)v_1^2 + \frac{1}{2}\lambda v_0^2 \\
v_1 \mu^2 &= \frac{1}{2}(15 a + 27 b)v^2v_1 + \frac{1}{2}(a+b)v_1^3 + \frac{1}{2}(2\alpha + \beta)v_0^2v_1 - \frac{3}{2} \beta v v_0^2\\
v \mu^2 &= (\frac{15}{2}a + \frac{7}{2}b)v^3 + (\frac{1}{2}a + \frac{9}{10}b)v_1^2v + (\alpha + \frac{3}{10}\beta)v_0^2 v - \frac{1}{10}\beta v_0^2v_1
\end{align}
As mentioned before, in the standard viewpoint the vevs are regarded as the output parameters. Therefore we should solve the minimum equations with respect to the vevs. However, doing so will in general result in a theory with no hierarchy. If both mass parameters $\mu^2$ and $\nu^2$ are \order{10^{14}} GeV then generally the vevs will also be of this scale. To ensure that the theory has a hierachy, the values of $v$ and $v_0$ will be imposed, and the minimum equations will be solved for three other parameters. Afterwards the fine-tuning necessary to obtain this hierarchy can be obtained by solving the system for the vevs.

To see whether all three vevs are really necessary, we can check the situation where $v_0 = 0$, so only the $\Phi$ field gets a vev.
In that case the minimum equations are given by:
\begin{align}
\mu^2 &= \frac{3}{2}(5 a + 9 b) v^2 + \frac{1}{2}(a+b)v_1^2 \\
\mu^2 &= \frac{15a + 7 b}{2}v^2 + \left(\frac{1}{2}a + \frac{9}{10}b\right)v_1^2
\end{align}
Solving this system for the two vevs shows that both vevs will be \order{\mu}, which makes sense since it is the only scale present in the system. The exact solutions show that $v_1 = 5 v$, so the two vevs will always be of the same order of magnitude and there is not even a possibility to fine-tune the parameters in order to obtain a hierarchy in the vevs. The $SU(5)$ symmetry will immediately be broken to $SU(3)_C \times U(1)_Q$, and the electroweak gauge bosons will be super heavy. This situation is clearly unwanted, and this option is therefore discarded. 

Finally, it is also possible to break $SU(5)$ directly to the Standard Model by setting $v = 0$ and having both $v_1$ and $v_0$ non-zero. However this situation again leads to the constraint $\beta = 0$, so this option is also discarded. Therefore we conclude that the only viable option is to have all three vevs $v$, $v_0$ and $v_1$ non-zero.

So we are forced to set both $v_1 \neq 0$ and $v_0 \neq 0 $ and impose the hierarchy that $v_0 \ll v$, with $v_0 \sim$ \order{100} GeV and $v \sim$ \order{10^{14}} GeV. However for $v_1$ it is not clear how large it should be.
Therefore we will use the minimum equations to figure out how large $v_1$ is. This is possible since we have three minimum equations, which we can solve for the two mass parameters and for $v_1$. Doing this gives rise to three separate solutions, with the result for $v_1$ for each of these solutions given by:
\begin{align}
1) \quad v_1 &= 5 v\\
2) \quad v_1 &= \frac{20 b v^2 + \beta v_0^2 - \sqrt{400 b^2 v^4 - 8 b \beta v^2 v_0^2 + \beta^2 v_0^4}}{8 b v} \\
3) \quad v_1 &= \frac{20 b v^2 + \beta v_0^2 + \sqrt{400 b^2 v^4 - 8 b \beta v^2 v_0^2 + \beta^2 v_0^4}}{8 b v}
\end{align}
We can approximate the second and third expression by expanding in small $v_0$. Depending on the sign of $b$, these expressions switch places, but the two results are:
\begin{align}
\quad v_1 &\approx \frac{3\beta v_0^2}{20 b v} \\
\quad v_1 &\approx 5v + \frac{\beta v_0^2}{10 b v}
\end{align}
So we see that in two out of the three cases $v_1$ will be of the order of the high scale, while in the other case $v_1$ is very small. We will show that using phenomenological constraints from the $\rho$ parameter, the solutions with large values for $v_1$ can be discarded.

The $\rho$ parameter is an electroweak observable that puts strong constraints on the Higgs sector. After decomposing all Higgs multiplets into Standard Model representations, the $\rho$ parameter is given by:
\begin{equation}
\rho = \frac{\sum_i[4T_i(T_i+1) - Y_i^2]v_i^2c_i}{\sum_i 2 Y_i^2 v_i^2},
\end{equation}
with $T_i$ the isospin, $Y_i$ the hypercharge, $v_i$ the vev and $c_i = 1/2$ $(1)$ for real (complex) representations. The experimental value is $\rho = 1.00039 \pm 0.00019$ \cite{PDG}, while the Standard Model (tree-level) result is $\rho = 1$, so only very small deviations are still allowed. Explicitly, the decompositions of $H$ and $\Phi$ are given by:
\begin{align}
5 &\rightarrow (3,1,-1/2) + (1,2,1/2)\\
24 &\rightarrow (8,1,0) + (1,3,0) + (1,1,0) + (3,2,-5/6q) + (3^*,2,5/6q)
\end{align}
The only contributions to $\rho$ arise from representations that get a vev. So the $(1,2,\frac{1}{2})$ of $H$ contributes (just like in the Standard Model) with vev $v_0$. The vev $v$ is part of the $(1,1,0)$ and has $T = 0$ and $Y = 0$, so it does not contribute to $\rho$. The only remaining contribution comes from $v_1$. It is part of the $SU(2)$ triplet field $(1,3,0)$. Therefore $\rho$ is given by:
\begin{align}
\rho &= \frac{\sum_i[4T_i(T_i+1) - Y_i^2]v_i^2c_i}{\sum_i 2 Y_i^2 v_i^2}\nn \\
%&= \frac{(4 \cdot 1/2 \cdot 3/2 -1)v_0^2 + (4 \cdot 1 \cdot 2-0)v_1^2 \cdot 1/2}{2 \cdot 1 \cdot v_0^2+2 \cdot 0 \cdot v_1^2} \nn\\
&= \frac{2v_0^2+4v_1^2}{2v_0^2} = 1 + 2\frac{v_1^2}{v_0^2}
\end{align}
So we see that to keep $\rho \approx 1$, we need to have $v_1 \ll v_0$. Therefore we can justify using only the $v_1 \sim \frac{v_0^2}{v}$ solution. This solution satisfies the bounds on $\rho$.

Now we can also justify our choice of imposing a $Z_2$ symmetry on the potential. It is possible to get a correct mass spectrum when setting $v_1 = 0$ by omitting the $Z_2$ symmetry \cite{Dorsner:2005ii}. In that case there are two additional terms in the potential: $\Lambda_1 \text{Tr}(\Phi^3)$ and $\Lambda_2 H^\dagger \Phi H$ with $\Lambda_1$ and $\Lambda_2$ couplings with mass dimension one. When setting $v_1 = 0$ the minimum equations fix $\Lambda_2 = 3\beta v$. This is independent of the inclusion of the $\Lambda_1$ term. In this case there is tree-level non-decoupling, meaning that there would be \order{1} contributions to the quartic scalar coupling of $H$ at low energy arising from the exchange of heavy $\Phi$ bosons due to the new $H^{\dagger}\Phi H$ interaction \cite{Li:1996mg}. Since this non-decoupling is undesired, we see that also in the case without $Z_2$ symmetry it is necessary to include $v_1$. Therefore we chose to impose the $Z_2$ symmetry in order to simplify the discussion and since this is the case most commonly discussed in the literature.

\section{Scalar masses}
\label{sec:ScalarMasses}
Now that we have a model where both $SU(5)$ breaking and EWSB are properly taken into account, with the hierarchy $v \gg v_0 \gg v_1$, we can take another look at the scalar masses. We will write these masses in terms of the vevs, since this most clearly illustrates the hierarchy in masses. Since we are interested in the doublet-triplet splitting problem we will focus on the masses of the components of $H$. But these components will mix with components of $\Phi$, so this mixing also needs to be taken into account.

First we look at the triplet components of $H$. The real scalar field $H_1$ mixes with $\phi_{11}$. After inserting the minimization conditions for $\mu^2$ and $\nu^2$ we get:
\begin{equation}
M^2(\phi_{11},H_1) = \begin{pmatrix}
\frac{1}{10 v}[-4b v v_1(v_1 + 5 v) + \beta v_0^2(2 v+v_1)] & -\frac{\beta v_0(v-v1)}{2\sqrt{2}} \\  
-\frac{\beta v_0(v-v1)}{2\sqrt{2}} & -\frac{1}{4}\beta (5 v^2 -6 v v_1+ v_1^2)
\end{pmatrix}.
\end{equation}
Knowing that $v_1 \sim v_0^2/v$, we can approximate these expressions, keeping only terms of \order{v_0^2} or larger:
\begin{equation}
M^2(\phi_{11},H_1) \approx \begin{pmatrix}
-2b v_1 v + \frac{1}{5}\beta v_0^2 & -\frac{\beta v_0v}{2\sqrt{2}} \\  
-\frac{\beta v_0 v}{2\sqrt{2}} & -\frac{1}{4}\beta (5 v^2 -6 v v_1)
\end{pmatrix}.
\end{equation}
So the $H$-$H$ component has mass of \order{v^2}, while the $\phi$-$\phi$ component has mass of \order{v_0^2}. After diagonalizing this mass matrix we get one massless mode and a mode with mass squared of \order{v^2}. This massless mode is a Goldstone boson arising from the breaking of the $SU(5)$ symmetry. The same result holds for the other components of the $H$ triplet. This is an exact result, it is not due to the approximation that was made.

For the components in the doublet we will discuss two cases: on the one hand we have the three components that do not get a vev, while on the other hand there is the component that does get a vev.

Each component that does not get a vev will mix with a component of $\Phi$. As an example we can look at the $H_7$ component, which mixes with $\phi_{21}$:
\begin{equation}
M^2(\phi_{21},H_7) = \begin{pmatrix}
\frac{1}{2}[20 b v^2 - \frac{4}{5}b v_1^2 + \frac{2}{5}\beta v_0^2 + \frac{v_1 v_0^2}{5v}\beta] & -\frac{3 v v _0}{2\sqrt{2}}\beta \\
-\frac{3 v v _0}{2\sqrt{2}}\beta & 3 v v_1 \beta
\end{pmatrix}
\end{equation}
We see that now the $\phi$-$\phi$ component has mass \order{v^2}, while the $H$-$H$ component is much lighter with mass \order{v_0^2}. After diagonalizing we again get a massless mode and a mode with mass \order{v^2}. This massless mode arises from the breaking of the electroweak symmetry. The $H_8$ and $H_{10}$ components behave in the same way.

To finalise our discussion on the masses of the scalars, we look at the $H_9$ component, which gets a vev in EWSB. This component mixes with $\phi_{23}$ and $\phi_{24}$:
\begin{align}
\begin{split}
&M^2(\phi_{23}, \phi_{24}, H_9) = \\
&\begin{pmatrix}
\frac{100 b v^3 + (10a + 6b)v v_1^2 + (2v + v_1)\beta v_0^2}{40 v} & -\frac{1}{8}\sqrt{\frac{3}{5}} (10a+18b v v_1 + \beta v_0^2) & -\frac{1}{4}v_0 (-3\beta v + 2 \alpha v_1 + \beta v_1) \\
-\frac{1}{8}\sqrt{\frac{3}{5}} (10a+18b v v_1 + \beta v_0^2) & \frac{(150 a + 70 b)v^3 + \beta v_0^2 v_1}{40 v} & \frac{1}{4}\sqrt{\frac{3}{5}}v_0 (10 \alpha v + 3 \beta v - \beta v_1) \\
-\frac{1}{4}v_0 (-3\beta v + 2 \alpha v_1 + \beta v_1) & \frac{1}{4}\sqrt{\frac{3}{5}}v_0 (10 \alpha v + 3 \beta v - \beta v_1) & \frac{\lambda v_0^2}{4}
\end{pmatrix}
\end{split}
\end{align}
After inserting the expression for $v_1$ and expanding in small $v_0$, we find that there is one eigenvalue of \order{v_0^2} and two eigenvalues of \order{v^2}.

All in all we find that after breaking $SU(5)$ to the Standard Model and then to $SU(3) \times U(1)$, we get 15 massless modes: 12 from breaking $SU(5)$ to the Standard Model and 3 more because of EWSB. The other scalars will be massive, with masses of \order{v^2}, except for one scalar which takes the role of the SM Higgs boson, with a mass of \order{v_0^2}. This is exactly as desired. Note that there are no scalars with a mass of \order{v_1^2}. 

\section{Amount of fine-tuning in the minimal $SU(5)$ GUT}
Now we have all the ingredients to quantitatively assess the amount of fine-tuning in the minimal $SU(5)$ GUT in the standard viewpoint. In this section we will compare the results of two different fine-tuning measures in this viewpoint.

The Dekens measure \cite{Dekens:2014ina} can be used to find fine-tuning present in the minimum equations. To apply this measure, one has to separate the set of parameters into two sets: the dependent parameters $q_j$ and the independent parameters $p_i$. The minimum equations relate the $q_j$ to the $p_i$. The amount of fine-tuning is then defined as:
\begin{equation}
\Delta_D = \max_{i,j} \Delta_D(p_i,q_j) = \max_{i,j} \left| \frac{p_i}{q_j} \frac{\partial q_j}{\partial p_i} \right|
\end{equation}

When the $q_j$ are polynomials in the $p_i$, this expression compares the size of each single contribution to the size of $q_j$. If $\Delta_D$ is large, there are contributions that are much larger than the size of $q_j$, which means that there needs to be a large cancellation between independent terms, implying fine-tuning.

In the original description of the Dekens measure, there is no prescription for which parameters should be used as the $q_j$. In the original paper, all possibilities are checked, and the set of $q_j$ which gives the maximum amount of fine-tuning is selected. But as argued in section \ref{sec:EWSB}, in the standard viewpoint, the vevs are regarded as output parameters. Even though the minimum equations might be solved for a different set of parameters, this is just a calculational tool used to consistently obtain a hierarchy in vevs. So the proper way to apply the Dekens measure in the standard viewpoint is to use the vevs as the $q_j$.

\begin{figure}
\centering
   \includegraphics[width=0.8\textwidth]{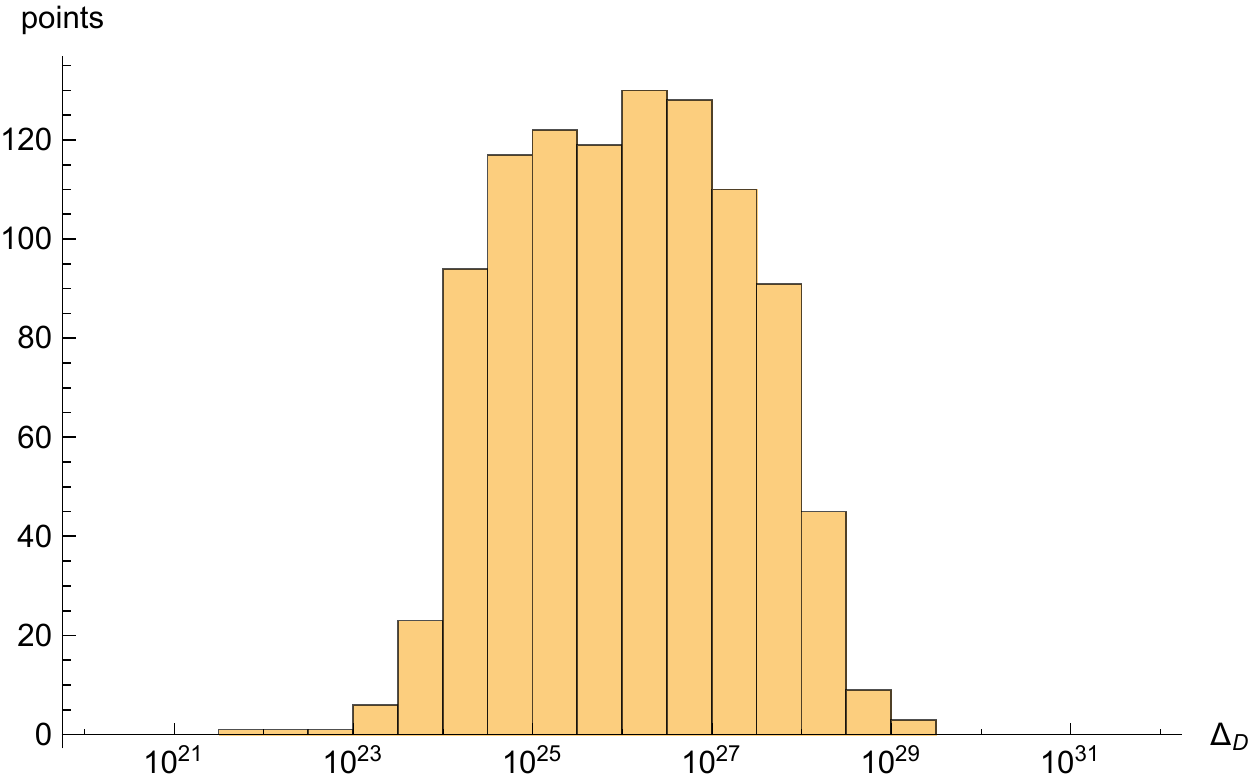}
   \caption{The result of calculating the Dekens measure in a minimal $SU(5)$ GUT using the vevs as dependent parameters. 500 points were sampled using the parameter ranges in Table \ref{tab:ranges}.}
   \label{fig:DekensStandard}
\end{figure}

While the minimum equations in our model are a complicated coupled system of equations in terms of $v$, $v_0$ and $v_1$, we can use the knowledge about the hierarchy to simplify the system. Writing $v_1 \sim v_0^2/v$, and keeping only terms in the minimum equations of $\mathcal{O}(v_0^2)$ and larger, we get:
\begin{align}
\nu^2 &= (15\alpha + \frac{9}{2}\beta)v^2 - 3\beta v v_1 + \frac{1}{2}\lambda v_0^2 \\
v_1 \mu^2 &= \frac{1}{2}(15 a + 27 b)v^2v_1 + \frac{1}{2}(2\alpha + \beta)v_0^2v_1 - \frac{3}{2} \beta v v_0^2\\
v \mu^2 &= (\frac{15}{2}a + \frac{7}{2}b)v^3 + (\alpha + \frac{3}{10}\beta)v_0^2 v
\end{align}
This system of equations can be solved analytically for the vevs, which means the Dekens measure can also be evaluated analytically. The analytical result is not very insightful however. It is easier to do a numerical study. We start from sampling the vevs $v$ and $v_0$ in the ranges $v = 10^{14} - 10^{16}$ GeV, $v_0 = 200-300$ GeV. The coupling constants are sampled in the range $-1$ to 1, taking into account some constraints to ensure the potential is bounded from below. The exact parameter ranges are shown in Table \ref{tab:ranges}. These values are used to determine $\mu^2$, $\nu^2$ and $v_1$. Then these values are used in the formula for the Dekens measure, where the mass parameters are now taken as input parameters. This gives the result shown in Figure \ref{fig:DekensStandard}.

\begin{table}
\centering
\begin{tabular}{ |c|c| } 
\hline
Parameter & Range \\
\hline
$v$ &  [$10^{14}$,$10^{16}$] GeV \\
$v_0$ & [200,300] GeV  \\
$\alpha$ & [-1,1]\\
$\beta, b, \lambda$ & [0,1]\\
$a$ & [-$b$,1]\\
\hline
\end{tabular}
\caption{Parameter ranges for calculating the Dekens measure in Figure \ref{fig:DekensStandard}, \ref{fig:BGstandard} and \ref{fig:DekensMeasure} .}
\label{tab:ranges}
\end{table}
We see that the Dekens measure consistently shows a large amount of fine-tuning  of the order $\frac{v^2}{v_0^2}$. This result agrees with results in the literature on the gauge hierarchy problem, see e.g. \cite{Mohapatra} %\cite{Gildener:1976ai,Natale:1982mt,Mohapatra}. 

The other fine-tuning measure we will investigate is the Barbieri-Giudice (BG) measure $\Delta_{BG}$ \cite{Barbieri:1987fn,Ellis:1986yg}, which determines the amount of fine-tuning in an observable with respect to the input parameters $p_i$ of the model. We will use the scalar masses as observables, so our expression for the BG measure is:
\begin{equation}
\Delta_{BG} = \max_{i,j}\left| \frac{p_i}{m_j^2}\frac{\partial m_j^2}{\partial p_i} \right|
\end{equation}

\begin{figure}
\centering
   \includegraphics[width=0.8\textwidth]{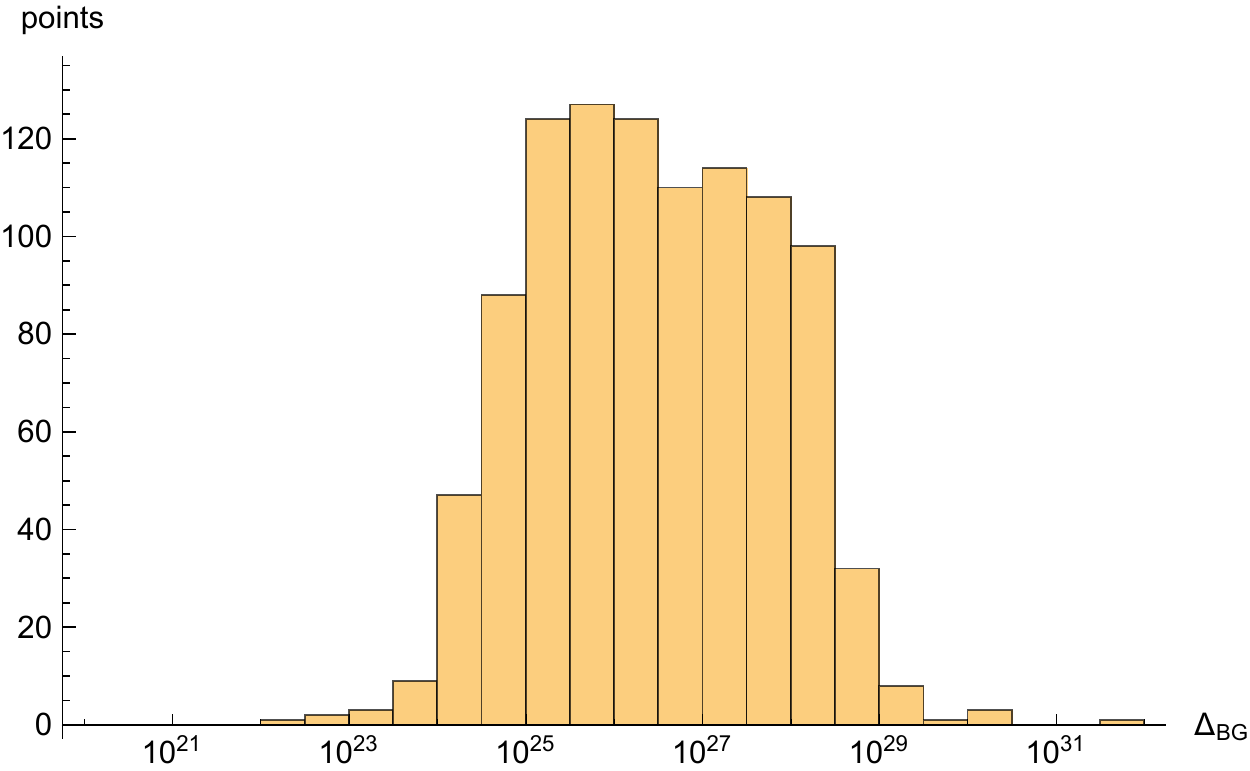}
   \caption{The result of calculating the BG measure in a minimal $SU(5)$ GUT in the standard viewpoint. 500 points were sampled using the parameter ranges in Table \ref{tab:ranges}.}
   \label{fig:BGstandard}
\end{figure}

Note that in order to calculate the fine-tuning, we need to write all the masses in terms of the input parameters. These expressions are too elaborate to show here, and calculating the derivatives has to be done numerically. The results are shown in Figure \ref{fig:BGstandard}.

We see that the results for the Dekens measure and the BG measure are extremely similar. This makes sense because we saw in the previous section that the masses are all proportional to a vev, so the results for the BG measure and the Dekens measure should agree up to some factors due to converting from the vevs to the masses.

This concludes the quantitative discussion of the doublet-triplet splitting in the standard viewpoint. Let us summarise the findings thus far. We have first considered how to break the $SU(5)$ symmetry, and later also the ways to subsequently break the electroweak symmetry. We described all possible ways to achieve this in the minimal $SU(5)$ model and reached the conclusion that in the standard viewpoint they should all be discarded because it requires fine-tuning of the parameters in the Lagrangian or it led to phenomenologically unacceptable side effects, like additional Goldstone bosons or large $\rho$ parameter corrections. It is clear that in none of the scenarios, a phenomenologically viable hierarchy of vevs will result from the generic theory with natural and perturbative parameters. The DTSP is an expression of this problem. But as we will show in the next section this problem is not unique to the $SU(5)$ GUT. Similar problems arise in simpler extensions of the SM with two coupled scalar sectors like the two Higgs doublet model (2HDM). 

\section{Hierarchies in the 2HDM}
The Higgs sector of the 2HDM is constructed by adding an additional doublet to the Standard Model Higgs sector \cite{Branco:2011iw, Boer:2019ipg}. The two doublets are defined as:
\begin{equation}
  \Phi_1 = \begin{pmatrix} \phi_1^+ \\ \phi_1^0 \\
\end{pmatrix},
\quad
\Phi_2 = \begin{pmatrix} \phi_2^+ \\ \phi_2^0 \\
\end{pmatrix}.
\end{equation}
In order to simplify our discussion, we demand $CP$ invariance and impose a $\mathbb{Z}_2$ symmetry on the potential. Under these constraints, the Higgs potential has the form:
\begin{align*}
V = -\mu_1^2 A - \mu_2^2 B +\lambda_1A^2 + \lambda_2B^2 +\lambda_3C^2+\lambda_4D^2+\lambda_5AB,
\end{align*}
where the invariants $A, B, C$ and $D$ are defined as:
\begin{align*}
  A = \Phi_1^{\dagger}\Phi_1, & \quad C = \frac{1}{2}\left(\Phi_1^{\dagger}\Phi_2 + \Phi_2^{\dagger}\Phi_1\right),\\
  B = \Phi_2^{\dagger}\Phi_2, & \quad D = \frac{1}{2i}\left(\Phi_1^{\dagger}\Phi_2 - \Phi_2^{\dagger}\Phi_1\right).
\end{align*}
The vacuum expectation values (vevs) for the two doublets are given by:
\begin{equation}
  \langle\Phi_1\rangle = \frac{1}{\sqrt{2}}\begin{pmatrix} 0 \\ v_1 \\
\end{pmatrix}
, \quad \langle\Phi_2\rangle = \frac{1}{\sqrt{2}}\begin{pmatrix} 0 \\ v_2 \\
\end{pmatrix}.
\end{equation}
In the standard phenomenological analyses of the 2HDM, the two vevs are required to satisfy the relation $v_1^2 + v_2^2 = v^2 = (246 \text{ GeV})^2$, which precludes a large hierarchy except for one of the vevs extremely close to zero. Here we will not impose this constraint in order to see if a large hierarchy of vevs can result from the general theory with parameters of \order{1}. 

Contrary to the $SU(5)$ GUT, we can analytically solve the minimum equations of the 2HDM for the vevs. Assuming that both $v_1$ and $v_2$ are non-zero, we find:
\begin{align}
\begin{split}
v_1^2 = \frac{\lambda_2\mu_1^2 - \lambda_+\mu_2^2}{\lambda_1\lambda_2 -\lambda_+^2},\\
v_2^2 = \frac{\lambda_1\mu_2^2 - \lambda_+\mu_1^2}{\lambda_1\lambda_2 -\lambda_+^2},
\end{split}
\label{eq:2HDM_vevs}
\end{align}
where $\lambda_+ \equiv  \frac{1}{2}(\lambda_3 + \lambda_5)$. To investigate how natural a hierarchy in vevs is, we will investigate two scenarios for the mass parameters $\mu_i^2$: the case where there is a large hierarchy in the mass parameters, say $\mu_1^2 \gg \mu_2^2$, and the case where they have very similar values, $\mu_1^2 \sim \mu_2^2$.

For a large hierarchy in the mass parameters, we see from Eq.\ (\ref{eq:2HDM_vevs}) that in the generic model both vevs will have values of \order{\mu_1}. So in general a hierarchy in the mass parameters will not translate into a hierarchy in the vevs. In that case there will not be any fine-tuning, neither in the BG measure nor in the Dekens measure with the vevs taken as output parameters. It is possible though to obtain a hierarchy in vevs in the special case that $\lambda_5 \approx - \lambda_3$ when there are actually two contributions of \order{\mu_1^2} to $v_2$ that cancel to give a much smaller value. The Dekens measure will show a large fine-tuning in this case and moreover, a slight variation in one of the parameters will lead to the hierarchy disappearing.

When the two mass parameters have similar values, again the generic model will have two vevs with similar values. There is no fine-tuning present then. But also in this case it is possible to obtain a hierarchy, although there is a different cancellation necessary to achieve this. When one of the two numerators in Eq.\ (\ref{eq:2HDM_vevs}) is very small, there will be a hierarchy in vevs. This again comes at the price of fine-tuning, due to a large cancellation between the contributions of the mass parameters.

So we can say in general that independent of the values one takes for the mass parameters, the generic theory will have no hierarchy in vevs. The reason for this is the coupling between the two scalar sectors. If the two sectors had been completely decoupled, a hierarchy in mass parameters would translate directly into a hierarchy in vevs. But due to the coupling $\lambda_+$, which relates the two sectors, any hierarchy in the $\mu_i^2$ gets nullified. This is also the key fine-tuning aspect of the DTSP in the $SU(5)$ GUT and a general problem when trying to achieve a hierarchy in vevs in BSM theories with extended scalar sectors.

The example with a cancellation between $\lambda_5$ and $\lambda_3$ shows that 
if the coupling between the scalar sectors is very small, a hierarchy in the mass parameters can still translate into a hierarchy in vevs.  We note that small $\lambda_+$ is also possible when $\lambda_3, \lambda_5 \ll 1$. If these small couplings are technically natural, for instance in case they arise effectively from an intermediate (portal) channel between the scalar sectors, the resulting hierarchy can be considered natural. But in the case of the $SU(5)$ GUT, and other GUTs, this does not work, because in order for the doublet and triplet fields to have masses of different scales, sufficiently large coupling between the $\Phi$ and $H$ fields is necessary, such that $\alpha$ and $\beta$ cannot both be small. This is another aspect of the DTSP in the $SU(5)$ GUT, which makes it harder to obtain a viable model in a natural way. All in all, we conclude that it is not possible to achieve a viable hierarchy in both vevs and scalar masses in a model like the minimal $SU(5)$ GUT with natural parameters and without fine-tuning.

The analysis of the $SU(5)$ GUT so far has assumed that the different symmetry breaking steps are independent, and that the scales at which the different symmetries break follow from the Lagrangian. But one may wonder whether one really can view the symmetry breaking pattern as independent steps and consider the vevs as pure output parameters? In the next section we will argue for a different viewpoint, where the hierarchy in vevs is used as input, and where all problems mentioned above are solved simultaneously.

\section{Alternative viewpoint on hierarchy and fine-tuning}
We showed in section \ref{sec:EWSB} and \ref{sec:ScalarMasses} that in order to have a hierarchy of triplet and doublet masses, without introducing additional Goldstone bosons or large corrections to the $\rho$ parameter, it is necessary to introduce a vev in the $\Phi$ field that breaks the electroweak symmetry. This means that already at the GUT breaking scale a (tiny) EWSB term is introduced, so the electroweak symmetry is actually never present below the $SU(5)$ breaking scale. This means that the two breaking mechanisms are not independent, since there is EWSB simultaneously with the breaking of the GUT. Therefore the theory at the GUT scale already has to ``know'' about the implementation of EWSB at low energy. To incorporate this, we advocate a viewpoint where besides the orientation of vevs ($\langle \Phi \rangle = \text{diag}(v, v, v, -\frac{3}{2}v -\frac{1}{2}v_1, -\frac{3}{2}v + \frac{1}{2}v_1)$ and $\langle H \rangle = \frac{1}{\sqrt{2}}(0,0,0,0,v_0)^T$ in this case) also the scale at which each breaking occurs is specified. In other words, the values of the vevs are imposed, or at least their relative values i.e.\ the hierarchy in vevs.

Note that it is standard practice to impose the symmetry breaking pattern by specifying the orientation of the vevs. This does not follow from the Lagrangian. Even in the case of just one vev this orientation must be imposed, since $\langle \Phi \rangle = \text{diag}(v, v, v, -\frac{3}{2}v, -\frac{3}{2}v)$ and $\langle \Phi \rangle = \text{diag}(v, v, v, v, -4v)$ amount to different breakings. Since one has to tie the $SU(5)$ symmetry breaking to the EWSB, a specific implementation is selected (imposed). If one would consider a generic theory with $SU(5)$ breaking at a high scale, followed at a lower scale by a {\it random} further breaking, then this will not be EWSB in general and certainly not EWSB at a much lower scale. By requesting the latter, one needs to fine-tune the parameters of the theory. But why not accept the fact that the second breaking is not random and explicitly select theories that have this breaking pattern and hierarchy of scales built in from the start? Instead of imposing symmetry $A$ at scale $a$ and expecting that generically symmetry $B$ will result at scale $b$, we simply accept that we need to restrict to theories that have symmetry $A$ at scale $a$ and symmetry $B$ at scale $b$. 
The presence of the required third vev $v_1$ in the $SU(5)$ case makes it unavoidable that the two breakings need to be adjusted to each other, but even in absence of $v_1$ the same line of reasoning can be made.

Just like imposing a symmetry will result in less free parameters, imposing a hierarchy on the vevs that break the symmetry will result in less free parameters. So in this viewpoint we are considering a specific subset of theories with less free parameters than the fully generic theory. We will adopt this viewpoint next and ask the more restricted question whether there is particular tuning needed within the subset of theories that lead to the desired hierarchy, or phrased differently, does the generic theory with the imposed constraints lead to the desired hierarchy for natural and perturbative parameters?

Since we propose to consider the vevs as input parameters, for illustration purposes we make them part of the Lagrangian via Lagrange multipliers, although this is not needed (the orientation of vevs is also not implemented in the Lagrangian). Using Lagrange multipliers (alternatively, one could consider constraint effective potentials \cite{ORaifeartaigh:1986axd}), we can impose the values of the vevs in the following way (for simplicity we restrict to the case $v_1 =0$):
\begin{equation}
\mathcal{L} = \mathcal{L}_0 + \Delta(\text{Tr}(\Phi^2) - \frac{15}{2}v^2) + \Delta_0(H^\dagger H - \frac{v_0^2}{2}),
\end{equation}
where $v$ and $v_0$ are to be understood as input parameters and $\Delta$ and $\Delta_0$ are Lagrange multipliers. These constraints are imposed in an $SU(5)$ invariant way\footnote{With $v_1 \neq 0$ the first constraint becomes Tr$\Phi^2 = \frac{15}{2}v^2 + \frac{1}{2}v_1^2$. So in order to fix both the value of $v$ and $v_1$, one needs to impose a second independent constraint on $\Phi$, e.g.\ $\Delta_1(\text{Tr }\Phi^4 - \frac{1}{8} \left(105 v^4 + 54 v^2 v_1^2+v_1^4\right))$.}. At the GUT breaking scale, we minimise the potential for both $\Phi$ and $\Delta$, which ensures that the GUT breaking minimum is located at the scale $v$. Then at the EWSB scale, we minimise with respect to $H$ and $\Delta_0$, ensuring the proper scale for $v_0$. In this way, the vevs are already present at the level of the Lagrangian.

Alternatively, we could write the constrained Lagrangian as:
\begin{equation}
\mathcal{L} = \mathcal{L}_0 + \Delta(x\text{Tr}(\Phi^2) - H^\dagger H)
\label{eq:constraintHierarchy}
\end{equation}
where $x$ quantifies the hierarchy in vevs: $x=\frac{v_0^2}{15v^2}$. In this case we minimise with respect to $\Delta$ only at the EWSB scale. The difference between the two approaches is that when we only impose the hierarchy, there is only one free parameter less, as opposed to two free parameters when we impose the value of both vevs. Since the doublet-triplet splitting problem only depends on the hierarchy, and not on the individual scales, it is enough to just add a constraint for the hierarchy. From now on we will work with this approach, but in practice there is no difference with imposing the values of the vevs, since the value of $v$ does not require fine-tuning, and imposing both $v$ and $x$ is equivalent to imposing the value of $v$ and $v_0$.

For any given $x$ value, the Lagrangian will thus contain one free parameter less, which is fully determined by the other parameters and the vevs. As our earlier analysis shows, variation of just a single parameter will result in a change of $x$ that generically will tend to be of \order {1}. In that sense small $x$ values are not stable under variation of a single parameter. But it can be stable when variation of one parameter is accompanied by variation of another one. The question then arises whether that should be considered fine-tuning. As argued in \cite{Boer:2019ipg} we would not call this fine-tuning if changes in one natural parameter requires changes in another one of exactly the same order.

This can then quantitatively be checked in an analysis of the Dekens measure, where the vevs are now considered input parameters. Now the question arises again what the proper way is to apply the Dekens measure: which choice should we make for the dependent parameters $q_j$?  As argued in \cite{Boer:2019ipg}, we will adopt the approach that we first solve the minimum equations for a certain set of parameters, and then apply the Dekens measure using these same parameters as the dependent parameters. If there is a set of parameters that leads to little or no fine-tuning, we argue that there is no fine-tuning necessary to maintain the hierarchy in vevs. 

\begin{figure}
\centering
   \includegraphics[width=0.8\textwidth]{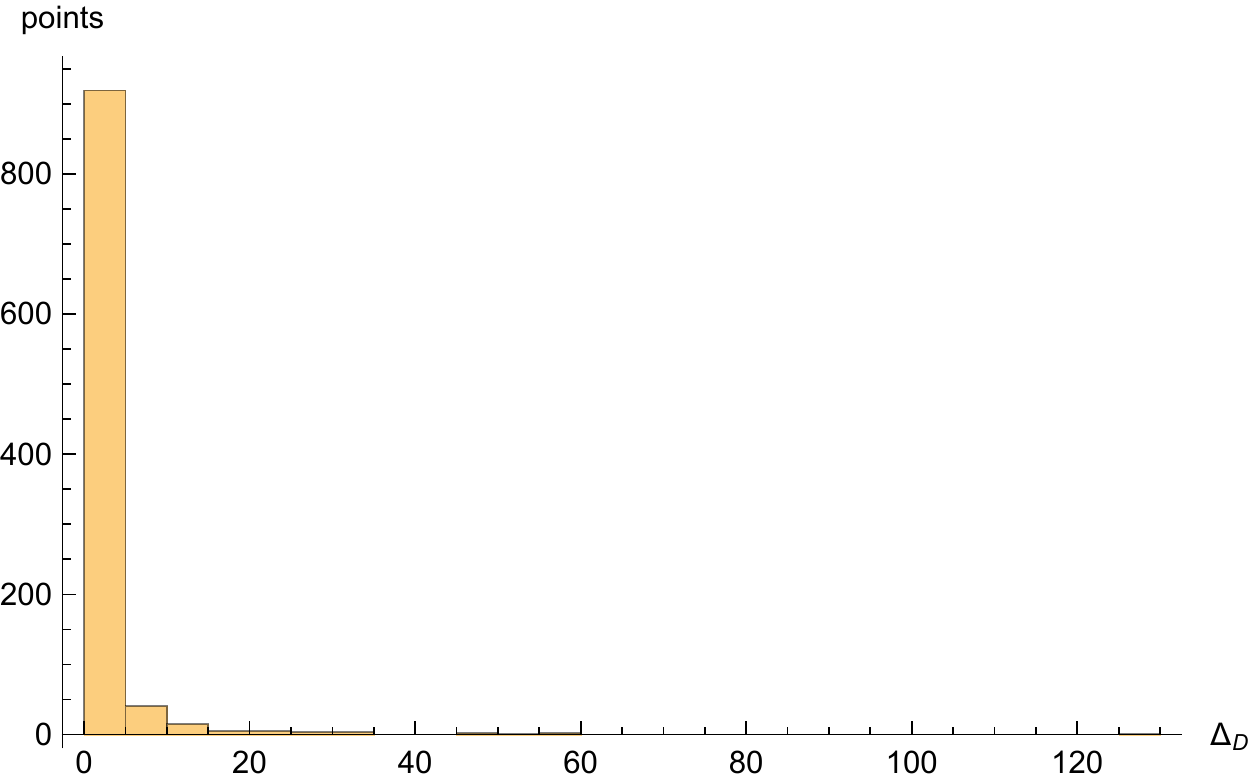}
   \caption{The result of calculating the Dekens measure in a minimal $SU(5)$ GUT using $\mu^2$, $\nu^2$ and $v_1$ as dependent parameters. 500 points were sampled using uniform distributions for the parameters with the ranges given in Table \ref{tab:ranges}.}
   \label{fig:DekensMeasure}
\end{figure}

For the minimal $SU(5)$ case, this means that we can take e.g.\ $\mu^2$, $\nu^2$ and $v_1$ as dependent parameters. Taking these parameters as the $q_j$, and imposing the hierarchy on the vevs as described above, the results for the Dekens measure are shown in Figure \ref{fig:DekensMeasure}. This shows that there is no fine-tuning present now, meaning that it is possible to maintain the hierarchy in vevs by adjusting parameters by the same order of magnitude.

The BG measure will also not show any fine-tuning in this scenario. All scalar masses are proportional to a vev, and the mass spectrum is exactly like it should be if the vevs have the proper values. So by imposing the vevs we ensure the correct mass spectrum and there will be no fine-tuning.

\section{Remarks on a large hierarchy in vevs}
In the previous section, we imposed the values of the vevs and their hierarchy and concluded that there is no fine-tuning necessary to maintain this hierarchy under variations of the remaining free parameters. But if one views the vevs as input parameters, that could even be part of the Lagrangian through constraint terms, then one may wonder if a large hierarchy of vevs should not be considered unnatural to begin with. The $x$ values in the numerical study of the previous section are of the order $10^{-23}$--$10^{-25}$. For such small values one might wish to have an explanation in terms of an enhanced symmetry when $x \rightarrow 0$, in accordance with the technical naturalness argument proposed by 't Hooft \cite{tHooft:1979rat}. Typically setting a vev to zero indeed enlarges the symmetry from a subgroup $H$ of a group $G$ to the full group $G$. However, in the case considered in this paper setting $v_0$ or $v_1$ to zero does not enhance the symmetry, unless one sets them both to zero. But even in that case one may wonder whether the symmetry is really enhanced, since the full group $G$ is still nonlinearly realised in the broken phase \cite{Coleman:1969sm}. In this sense the full symmetry is still present in the underlying description of the theory and no enhanced symmetry ever arises if vevs are set to zero. If the symmetry argument can then not be applied to vevs, it is also not required to apply 't Hooft's naturalness criterion to them. This is the viewpoint we advocate, because the only alternatives are to only accept hierarchies of \order{1} or to view the vevs as output parameters, which effectively boils down to the same conclusion when one starts from the most general unconstrained theory with natural and perturbative parameters. Imposing the hierarchy as part of the theory, on top of imposing the symmetry breaking pattern, would then be a way to obtain a viable model.

\section{Conclusion}
We have studied the symmetry breaking pattern in the minimal $SU(5)$ GUT, and confirmed in line with the DTSP that it is not possible to obtain the desired hierarchy of vevs without fine-tuning in a theory with all natural and perturbative parameters. We also showed that this is actually a general problem of coupled scalar sectors: the coupling between the two sectors will prevent a hierarchy in mass parameters to translate into a hierarchy in vevs. We advocate for a different viewpoint, since it is not possible to view the breaking of the GUT symmetry separate from the EWSB, which implies that the theory at the GUT scale has to be adjusted to the low-energy physics by imposing the orientations of the vevs. We argue that also the hierarchy needs to be imposed such that the vevs are not considered output parameters anymore. For illustration purposes we incorporated this idea by imposing the hierarchy in vevs by using a Lagrangian multiplier. We find that the generic theory with a large hierarchy in vevs will not need any fine-tuning of the remaining free parameters in order to produce the correct mass spectrum. This means that changes in one natural parameter can be countered by changes in another natural parameter of the same order of magnitude. Furthermore, we argue that a large hierarchy in vevs need not be considered unnatural to begin with, even though a large hierarchy will in general not arise from a generic unconstrained theory (and even if it does, a small change in the parameters would remove the hierarchy again). Finally, we emphasise once more that our discussion only applies to the classical level and does not shed any light on issues due to loop corrections.

\textbf{Acknowledgements}
We thank Ilja Dor\v{s}ner and Sybrand Zeinstra for useful feedback on the first version of this paper. This work has been financially supported by the NWO programme ``Higgs as a probe and portal".

\end{document}